\documentclass[preprint]{aastex}

\begin{document}
\parskip=0pt

\let\deg=\arcdeg

\def\ie{{\it i.e.,\ }}
\def\eg{{\it e.g.,\ }}
\def\qv{{\it q.v.,\ }}
\def\cf{{\it cf.\ }}
\def\etaln{{\it et al.}}
\def\etal{{\it et al.\ }}
\def\gtrapprox{\;\lower 0.5ex\hbox{$\buildrel >
    \over \sim\ $}}             
\def\lessapprox{\;\lower 0.5ex\hbox{$\buildrel < \over \sim\ $}}
\def\msol{\ifmmode {\>M_\odot}\else {$M_\odot$}\fi}
\def\pyr{\ifmmode {\>{\rm\ yr}^{-1}}\else {yr$^{-1}$}\fi}
\def\psec{\ifmmode {\>{\rm\ s}^{-1}}\else {s$^{-1}$}\fi}
\def\kms{\ifmmode {\>{\rm km\ s}^{-1}}\else {km s$^{-1}$}\fi}
\def\psqcm{\ifmmode {\>{\rm cm}^{-2}}\else {cm$^{-2}$}\fi}
\def\pcubcm{\ifmmode {\>{\rm cm}^{-3}}\else {cm$^{-3}$}\fi}
\def\phoflux{\ifmmode{{\rm phot\ cm}^{-2}{\rm\ s}^{-1}}\else {phot
    cm$^{-2}$ s$^{-1}$}\fi}
\def\intensity{\ifmmode{{\rm W\ m}^{-2}{\rm\ Hz}^{-1}{\rm\ sr}^{-1}}
      \else {W m$^{-2}$ Hz$^{-1}$ sr$^{-1}$}\fi}

\def\be{\begin{equation}}
\def\ee{\end{equation}}
\def\bea{\begin{eqnarray}}
\def\eea{\end{eqnarray}}

\def\pma#1 #2 {\lower 0.3ex\hbox{$\buildrel {\scriptscriptstyle + #1} \over 
{\scriptscriptstyle - #2}$}\ }

\title{A Fluctuation Analysis of the Bolocam 1.1mm Lockman Hole Survey}
\author{Philip R.~Maloney\altaffilmark{1,3}, Jason
  Glenn\altaffilmark{1,2}, James E.~Aguirre\altaffilmark{1}, Sunil
  R.~Golwala\altaffilmark{4}, G.~T.~Laurent\altaffilmark{1},
P.~A.~R.~Ade\altaffilmark{5}, 
J.~J.~Bock\altaffilmark{4,6},
S.~F.~Edgington\altaffilmark{4},
A.~Goldin\altaffilmark{6},
D.~Haig\altaffilmark{5},
A.~E.~Lange\altaffilmark{4},
P.~D.~Mauskopf\altaffilmark{5},
H.~Nguyen\altaffilmark{6},
P.~Rossinot\altaffilmark{4},
J.~Sayers\altaffilmark{4}, \&
P.~Stover\altaffilmark{1}}

\altaffiltext{1}{Center for Astrophysics and Space Astronomy,
       University of Colorado, Boulder, CO 80309-0389}
\altaffiltext{2}{Dept. of Astrophysical and Planetary Sciences,
       University of Colorado, Boulder, CO 80309-0593}
\altaffiltext{3}{maloney@casa.colorado.edu}
\altaffiltext{4}{California Institute of Technology, 1200
       E.~California Blvd., MC 59-33, Pasadena, CA 91125}
\altaffiltext{5}{Physics and Astronomy, Cardiff University, 5 The Parade,
Cardiff CF24 3YB, Wales, UK}
\altaffiltext{6}{Jet Propulsion Laboratory, California Institute of
Technology, 4800 Oak Grove Drive, Pasadena, CA 91109}

\begin{abstract}
We perform a fluctuation analysis of the 1.1mm Bolocam Lockman Hole
Survey, which covers 324 arcmin$^2$ to a very uniform point
source-filtered RMS noise level of $\sigma\simeq 1.4$ mJy beam$^{-1}$.
The fluctuation analysis has the significant advantage of utilizing
all of the available data, since no extraction of sources is
performed: direct comparison is made between the observed pixel flux
density distribution ($P(D)$) and the theoretical distributions for a
broad range of power-law number-count models, $n(S)=n_o
S^{-\delta}$. We constrain the number counts in the 1-10 mJy range,
and derive significantly tighter constraints than in previous work:
the power-law index $\delta=2.7$\pma 0.18 0.15 , while the amplitude
is $n_o = 1595$\pma 85 238 mJy$^{-1}$ deg$^{-2}$, or $N(>1{\;\rm
mJy})= 940$\pma 50 140 deg$^{-2}$ (95\% confidence). At flux densities
above 4 mJy, where a valid comparison can be made, our results agree
extremely well with those derived from the extracted source number
counts by \citet{laur05}: the best-fitting differential slope is
somewhat shallower ($\delta=2.7$ versus $3.2$), but well within the
68\% confidence limit, and the amplitudes (number of sources per
square degree) agree to 10\%. At 1 mJy, however (the limit of the
$P(D)$ analysis), the shallower slope derived here implies a
substantially smaller amplitude for the integral number counts than
extrapolation from above 4 mJy would predict. Our derived
normalization is about 2.5 times smaller than determined by MAMBO at
1.2mm \citep{greve04}. However, the uncertainty in the normalization
for both data sets is dominated by the systematic (\ie absolute flux
calibration) rather than statistical errors; within these
uncertainties, our results are in agreement. Our best-fit amplitude
at 1 mJy is also about a factor of three below the prediction of
\citet{blain02}, but we are in agreement above a few mJy. We estimate
that about 7\% of the 1.1mm background has been resolved at 1 mJy.

\end{abstract}


\section{Introduction}
The study of background radiation fields - the integrated contribution
from objects over all redshifts - at different wavelengths has
provided valuable constraints on the history of the universe.  Since
different wavelengths are dominated by different classes of objects
and, in effect, by different physical processes, it is possible to
place potentially powerful constraints on the history (i.e., the
luminosity function and redshift distribution) of a chosen class of
object by the choice of waveband \citep[\eg][]{hd01,kash05}.

The detection of the cosmic infrared background (CIB) by the COBE
satellite \citep{pug96,fix98} offered a new view of galaxy
evolution. The surprisingly large amount of energy in the CIB indicates
that the total luminosity from thermal dust emission is comparable to
or exceeds the integrated UV/optical energy output of galaxies
\citep{guid97}. The only plausible sources of this luminosity are
dusty star-forming galaxies, or dust-enshrouded AGN.

The discovery of the CIB was rapidly followed by deep surveys, both
from the ground (with SCUBA, Bolocam, and Mambo at $850\mu$m, 1.1mm,
and 1.3mm, respectively) and from space, using ISO (at 15, 90, and
$170\mu m$). The high number counts (compared to no-evolution or
moderate-evolution models for the infrared galaxy population) found in
all these surveys imply that strong evolution of the source
populations must have occurred \citep[\eg][and references
therein]{scott02,lag03}, implying that these observations probe a
major epoch in the history of the universe. Comparison of the number
counts with the observed CIB indicates that only a small fraction
($\lessapprox 10\%$) of the background has been resolved in the
far-infrared. A similar fraction (10-20\%) is resolved in the
submillimeter in blind sky surveys, but by taking advantage of
gravitational lensing, it is possible to go deeper, and the SCUBA Lens
Survey has resolved $\sim 60\%$ of the background at $850\mu$m
\citep{smail}.

The deep Bolocam survey of the Lockman Hole \citep{laur05} covered 324
square arcminutes to a very uniform RMS noise level $\sigma\simeq 1.8$
mJy; after optimally filtering for point sources, the RMS in the
uniform coverage area is $\simeq 1.4$ mJy \citep[see][for
details]{laur05}. In that paper, the source number counts were
determined by first extracting point source candidates, then
performing extensive simulations and tests to establish the robustness
of the source candidates, and to estimate the number of false
detections and the effects of bias and completeness on the derived
number counts. The effects of Eddington bias - the upward bias of
source flux densities by noise fluctuations - in particular, are
substantial, such that most of the detected sources actually have flux
densities that lie below the formal $3\sigma$ detection limit.

An alternative approach, which avoids the requirement of identifying
and extracting point sources, is to analyze the distribution of pixels
in the map as a function of flux density \citep{scheuer57,
con74}. This type of fluctuation analysis is frequently referred to as
a ``$P(D)$'' analysis (the ``$D$'' denoting ``deflection''), following
early radio astronomical terminology. The great advantage of this
method is that all of the data are used, thereby making it possible to
derive information about sources with flux densities lying below the
formal detection limit, whereas any sensible point source extraction
technique must implement a minimum S/N ratio for acceptable
sources. In principle, a fluctuation analysis provides information on
the source distribution down to the flux density level at which there
is approximately 1 source per beam \citep{scheuer74}
provided the noise level is sufficiently
low. A meaningful fluctuation analysis does require, however, that the
noise in the map is very well-understood and characterized.

This latter requirement is amply satisfied by the Bolocam Lockman Hole
data. Since accurate determination of systematic effects on the number
counts (\eg contamination by spurious sources, flux boosting by
Eddington bias) requires a thorough understanding of the noise in the
data set, \citet{laur05} invested considerable effort in
characterizing the remaining noise in the Lockman Hole map after
cleaning and sky subtraction. In brief, multiple realizations of
jackknife maps were constructed by randomly selecting 50\% of the data
and co-adding it into a map, doing the same with the other half of the
data, and then differencing the two maps. Any sources, which should be
coherent over multiple observations, will be removed from the
jackknife map. Hence the jackknife maps can be used to determine the
actual power spectral density (PSD) of the noise in the Lockman Hole
map, independent of the signal contribution. Only the uniform coverage
region, in which the rms in the integration time per pixel is no more
than 12\% (implying that the rms noise variation is no more than 6\%),
has been used in the analysis; the map has been corrected for coverage
variations prior to construction of the jackknife maps. Hence the
noise in the map is both very uniform and well-characterized.

In this paper we present a fluctuation analysis of the
Lockman Hole observations. 
In \S 2 we present the Bolocam data and describe the method of
analysis and present the results, while in \S 3 we compare the results
to those derived from the number counts by \citet{laur05}. Section 4
discusses our results and briefly comments on the implications for
further deep mm-wavelength surveys.


\section{$P(D)$ analysis of the Bolocam Lockman Hole Observations}
The extremely uniform and well-characterized noise in the Bolocam map
of the Lockman Hole makes this data set very well-suited for a
fluctuation analysis. Since this technique is completely independent
of the number count analysis in \citet{laur05}, which relied on
extraction of point sources, it is worth revisiting these data. With
an RMS noise level of 1.4 mJy in the optimally-filtered map, we expect
the fluctuation analysis to probe the number count distribution to the
$S\sim 1$ mJy level; at significantly lower flux densities the noise
will completely dominate over the signal in the map. The filtering of
the map will not affect the fluctuation analysis. Optimal filtering is
the act of convolving with a kernel that is optimal for point source
extraction (based on a frequency space signal to noise
weighting). Such a convolution is a linear mathematical operation, and
hence affects point sources of all fluxes in the same way. In
practice, the filtering kernel looks mostly like a gaussian (hence
removing sub-beam sized noise) but has a low-frequency roll-off to
minimize the contribution of $1/f$ noise. The result of
optimally-filtering the map for point sources is thus to improve the
signal to noise of sources the size of the PSF, independent of signal
strength. No physical sources can be smaller than the PSF, even if
they are too faint to be detected. Hence filtering the map reduces the
noise level without affecting the underlying source count
distribution (at the expense of worsened angular resolution). An
example of the effects of optimal filtering on a pixel distribution is
is presented in the Appendix.

Since the goal of this analysis is to probe the distribution of
high-redshift galaxies, it is imperative that we show that the Bolocam
Lockman Hole observations are not significantly contaminated by other
signals. There are three potential important sources: primary and
secondary cosmic microwave background (CMB) fluctuations, and Galactic
dust foreground emission.

The CMB primary and secondary (thermal Sunyaev-Zeldovich) spectra are
generally given in the form
\be
{\cal C}_l={l(l+1)\over 2\pi} C_l
\ee
where the $C_l$ are the squares of the individual multipole
amplitudes, given as $(\Delta T/T)^2$. The contribution from such a
background to the rms flux density in a given experiment is given by
summing
\be
{2 l+1\over 4\pi}C_l W_l^2
\ee
over all $l$, where $W_l$ is the window function that describes the
response of the experiment to power at a given multipole. For Bolocam,
$W_l$ can be split into two pieces, $W_l=B_l F_l$, where $B_l$
is the window function of the beam and $F_l$ is an effective spatial
filter which depends on the scan strategy and the sky subtraction and
cleaning algorithms. For a Gaussian beam, the beam window function can
be closely approximated by
\be
B_l=e^{-l(l+1)\sigma_b^2/2}
\ee
where $\sigma_b$ is the dispersion of the Gaussian
\citep{white92}. The effective spatial filter for a given instrument
can vary from one data set to another as the scan strategy is
changed. For the Lockman Hole map, $F_l$ has been determined
empirically by processing white noise maps of the same size as the map
through the data reduction pipeline. A very good fit is provided by
\be
F_l = A\left[1-{1\over 2\pi}{\tan^{-1}(2\pi l/l_o)\over
  (l/l_o)}\right]
\ee
with $A=0.93$ and $l_o=4393$ (which corresponds to a spatial frequency
$f_o=0.20$ arcmin$^{-1}$).

We have used the current best-fit model to the observed CMB
anisotropies \citep[\eg][]{stomp01} and model predictions for the S-Z
power spectrum \citep{zhang02,bond05} to estimate that the RMS
contributions are $\Delta S_{CMB}\simeq\Delta S_{SZ}\simeq 0.24$
mJy. These values should be compared to the RMS of 1.8 mJy in the
Lockman Hole map prior to optimal filtering. The near-identical values
for the CMB and S-Z signals are coincidental, as the spatial filtering
resulting from sky subtraction and cleaning produces a very large
reduction in the CMB contribution: without this filtering, the CMB
contribution to the rms would be 1.8 mJy.  

We have not explicitly calculated the Galactic dust contribution to
the Lockman Hole map. However, \citet{masi01} have analyzed the dust
contribution to BOOMERANG maps and demonstrate that for regions with
column densities as low as the Lockman Hole, the dust contribution is
negligible compared to the CMB at 275 GHz.

The measured $P(D)$ of an astronomical map is simply the pixel
distribution: the number of pixels in each flux density bin, summed
over the entire map. No flux density thresholds are imposed, and the
entire uniform coverage region is included. Because of the faintness
of the sources in the map, we have \citep[as in][]{laur05} taken
advantage of our knowledge of the Bolocam beam shape to optimally
filter the map for point sources. This increases the effective beam
size, which increases the amount of source confusion (see \S 4);
however, the improved S/N of the sources (since optimal filtering
reduces the noise level in the map) more than compensates for the
worsening of source confusion. Figure 1 displays the cleaned,
optimally-filtered Lockman Hole map (this is the same map presented by
Laurent \etaln), while Figure 2 shows the observed $P(D)$
distribution. A total of 66 bins were used, as a compromise between
resolution of the distribution and the number of pixels/bin (see
below). Note also that the mean has been subtracted from the map, so
the peak of the distribution is nearly (but not precisely, due to the
presence of sources in the map) at zero.

We need to construct theoretical $P(D)$ distributions to compare with
the observed distribution, for a given choice of source model (\ie
number of sources per unit area as a function of flux density). This
can be done directly for an assumed number count distribution
\citep[see \eg][]{ti04} provided the beam shape and the noise are well
characterized, but there are two complications for the Lockman Hole
data: the shape of the optimal filter is not analytic, and 
even within the uniform coverage area, there are fluctuations in the
number of observations per pixel, which translate to variations in the
noise level (at about the 6\% level, prior to optimal filtering); see
Figure 1 of \citet{laur05}. 

We have therefore taken an alternate approach, in which we construct
simulated maps from which we calculate the $P(D)$ distribution
directly, just as for the actual data. The source distribution in all
cases is a featureless power-law, with differential number count
distribution
\be
n(S)=n_o S^{-\delta}\ {\rm mJy^{-1}\ deg^{-2}}
\ee
where $S$ is the source peak flux density (in mJy). For all of the
simulations the flux density range used was 0.1 to 10 mJy; the lower
limit is small enough compared to the RMS noise that varying it has no
significant effect on the results (decreasing it merely increases the
DC level of the map, which is set to zero in any case), while the
upper limit simply ensures that there are no sources much brighter
than any present in the Lockman Hole. (In fact, because of the
steepness of most of the source distributions considered, the results
are generally independent of the choice of maximum flux, for any value
larger than this.) The sources are added to a blank map with peak flux
densities randomly drawn from the above power-law, as gaussian sources
with size matched to the pointing-smeared Bolocam beam FWHM of
$36.7''$. The simulated map pixels are fixed at $10''\times 10''$, as
in the real map shown in Figure 1. The sources are uniformly randomly
placed \citep[\ie zero spatial correlation; note that strong
clustering of the sources will affect the resulting $P(D)$
distribution:][]{barc92,ti04} over an area larger than the final map,
so that sources can fall only partially within the map.

In order to match the Bolocam observations, the noise is calculated
using the measured PSD  from the Lockman jackknife maps, which contain
no source signal \citep{laur05}: a white noise realization is
constructed in Fourier space, then multiplied by the jackknife PSD and
normalized to produce the correct noise RMS \citep[see discussion in
\S5.2 of][]{laur05}. In other words, the simulated noise map is
constructed so that it has, on average, both the same RMS and the same
power as a function of spatial frequency as the actual noise in the
Lockman Hole map. The resulting noise map is then corrected for the
coverage variations, and added to the source map. Although the PSD of
the same jackknife map was used for all of the simulations, we
verified that the difference between the PSDs of different jackknife
maps is no larger than the differences produced by using independent
white noise realizations multiplied by the same PSD, the procedure
used here\footnote{Although the PSD, being the square of the Fourier
transform of the map, does not preserve phase information, we do not
expect this to have significant effect on the $P(D)$
analysis. Generating a map from its PSD is equivalent to taking the
original map and re-arranging its pixels while preserving the
amplitude of its FFT. Unless the map is dominated by such regular
structure that phase cancellation effects are important, this
rearrangement will have no significant impact on the pixel
distribution.}.

Since there are only about 12,000 pixels in the good coverage region,
the effect of shot noise (in both the source and noise contributions)
in the simulations is substantial. To eliminate this as a source of
uncertainty, we initially generated between 50 and 150 independent
realizations for each choice of source power-law: in each realization
the sources are randomly drawn from the specified power-law, while a
random realization of the noise map is generated as described
above. The pixel flux density distribution for each of these maps is
then calculated, and these are averaged together to produce the
theoretical $P(D)$.

To analyze the fluctuations in the real data, we have carried out a
maximum likelihood analysis to compare the observed pixel distribution
with the predicted $P(D)$ from the simulations for a broad range of
power-law models. As discussed by \citet{fried}, as long as we ensure
that the number of pixels in each bin is large, so that the Poisson
distribution of the number of pixels per bin is closely approximated
by a Gaussian, and that there are negligible correlations between
pixel bins, then maximizing the likelihood function is equivalent to
minimizing 
\be
Q^2=\sum_{i=1}^{N_{bins}}\left({{p_i-\mu_i}\over\sqrt\mu_i}\right)^2
\ee 
where $p_i$ is the number of pixels within flux bin $i$ in the Lockman
Hole data and $\mu_i$ is the expected number of pixels in the bin as
predicted by the assumed noise-convolved model. In other words, if the
above assumptions are satisfied, then $Q^2$ is a good approximation to
$\chi^2$, and minimizing $Q^2$ is equivalent to maximum likelihood
estimation. The choice of 66 bins over the range $-4.5$ to $+4.8$ mJy
was made as a compromise between resolving the distribution and
keeping the number of pixels per bin large enough so that the above
assumption is reasonably well satisfied over most of the range. In the
actual calculation of $Q^2$, any bins with fewer than 10 pixels in
either the observed or predicted $P(D)$ were not included; the
smallest number of bins actually used was 56, and was typically 58 or
59\footnote{Note that since we are discarding some data - always, in
fact, the most extremal bins - we expect that the resulting confidence
regions for the model parameters will be more conservative than if we
had discarded no data. If we raise the minimum pixels per bin
threshold to 15 or 20, the size of the confidence regions increase, as
expected; the location of the minimum is unaffected.}. 
Both the best-fit values and the errors on the derived number-count
parameters were derived by directly mapping out $\chi^2$ space by
variation of the model parameters. Once the general shape of the
$\chi^2$ distribution was established, we ran a new set of models
covering the interesting region of parameter space, with 200
realizations of each model.


In Figure 3 (left-hand panel) we show the Lockman $P(D)$ from Figure 2
again, overplotted with the theoretical $P(D)$ produced by averaging
100 realizations of a noise-only simulation. The obvious discrepancies
between the two distributions - the higher peak and narrower width of
the noise-only $P(D)$ - are simply a reflection of the presence of
signal in the Lockman map. In the right-hand panel, we plot the
Lockman $P(D)$ again, now overplotted with the simulated $P(D)$ for
the best-fitting model, which has $\delta=2.7$ and $n_o=1595$
mJy$^{-1}$ deg$^{-2}$. The reduced chi-square for this model is
$\chi_\nu^2=0.90$ ($\chi^2=51.5$ for 57 degrees of freedom); for
reference, $\chi_\nu^2=9.3$ for the noise-only realization.

The number of sources per beam at the 1 mJy level is only about 0.1;
it is the noise level in the map, rather than the density of sources,
that limits the flux density we can probe with the fluctuation
analysis. Our best-fit model implies that Bolocam would reach the $\sim 1$
source per beam level at $S_\nu\approx 0.3$ mJy. We also note that,
although we could in principle have fit a more complicated model (\eg
a broken power-law) to the data, the goodness of fit of the
best-fitting model is quite high, with a 68\% probability (for 59 bins
- 2 model parameters = 57 degrees of freedom) that $\chi^2$ would
exceed this value by chance. Hence there is no compelling reason for
fitting a more complex model.

In Figure 4 we show the $\chi^2$ map of the power-law model
parameter space. The abscissa is the power-law index, $\delta$; for
the ordinate we have used the normalization $N(> 1{\;\rm mJy})$, the
number of sources per square degree with peak flux density greater
than or equal to 1 mJy (\ie the integral number count
distribution). This is related to $n_o$ by
\be
n_o=(\delta-1)N(>S) S^{\delta -1}
\ee
where $S$ is the peak flux density. The contours are $\Delta\chi^2$
for joint confidence limits on $\delta$ and $N$ of $68\%$, $95.4\%$,
and $99.7\%$, labeled with $1$ for $\Delta\chi^2=2.3$, $2$ for
$\Delta\chi^2=6.17$, and $3$ for $\Delta\chi^2=11.8$. The location of
the minimum is marked by a cross, with $N(> 1{\;\rm mJy}) \simeq 940$
deg$^{-2}$. To further reduce the noise, the contours have been mildly
smoothed by re-gridding via Delaunay triangulation. The 95\%
confidence limits on $\delta$ and $N$ (marginalizing over $N$ and
$\delta$, respectively) are $\delta=2.7$\pma 0.18 0.15 , $N=940$\pma
50 140 deg$^{-2}$, respectively. The equivalent confidence region for
$n_o$ is $n_o=1595$\pma 85 238 mJy$^{-1}$ deg$^{-2}$.

\section{Comparison with the Lockman Hole Point Source Results}
As noted above, \citet{laur05} determined the best-fitting number
counts in the observed Lockman Hole region by identifying point
sources, carrying out extensive simulations to determine the effects
of flux bias and completeness, then performing a maximum likelihood
analysis to constrain the allowed number counts (assumed to be a
power-law). Since only source candidates above $3\sigma$ were
considered, this analysis only provides information on the number
counts above an observed flux density $S_\nu\approx 4$ mJy. Hence the
only fair way to compare the results from this paper with the
\citet{laur05} results is to convert the $P(D)$ number count
constraints from 1 mJy to 4 mJy. This requires scaling the
normalization by a factor of $4^{1-\delta}$ (see equation [7]
above). We also need to rescale the \citet{laur05} results, since
their normalization parameter $A$ (chosen to reduce the degeneracy
between $n_o$ and $\delta$ when only a small range in flux density is
available) is related to the integral number counts by $N(> 4\;{\rm
mJy})=4A/(\delta-1)$ for $\delta > 1$.

In Figure 5 we show the re-scaled $P(D)$ results (note that the effect
of this correction is to suppress (in $N$) the higher-$\delta$ part of
the distribution and magnify the low-$\delta$ end, compared to the 1
mJy counts shown in Figure 4), with the same contours as in Figure
4. The minimum is located at $N(> 4\;{\rm mJy})\simeq 88.9$
deg$^{-2}$. Also plotted, in red, are the joint $68\%$ $N$-$\delta$
contour for the point source-derived number counts and the location of
the minimum (also in red) at $N(> 4\;{\rm mJy})\simeq 96.3$
deg$^{-2}$. Clearly, the results from both analyses are in good
agreement with one another: the $P(D)$ minimum lies within the
68\% confidence contour from the number-count analysis of \citet{laur05}. 
The 4 mJy normalizations differ by less than $10\%$. The smaller value
of the power-law index preferred by the fluctuation analysis is
undoubtedly a consequence of the greater dynamic range in flux density
that is included, since very few data are discarded. Because of the
steepness of the allowed power-laws ($\delta\sim 2.5$ - 3), a factor
of four decrease in flux density results in an order of magnitude
increase in the integrated number of sources, which accounts for the
much smaller errors on $N$ and $\delta$ produced by the fluctuation
analysis. 

In Figure 6 and 7 we show our derived number counts together with
the \citet{laur05} number counts, as well as all of the other directly
comparable observed and theoretical number counts (discussed
below). Figure 6 plots the differential number counts, while 7 shows
the cumulative number counts. The differential counts are much to be
preferred since they require no assumptions about the source
distribution at fluxes higher than have been observed; furthermore,
the integral counts tend to obscure any structure in the number count
distribution. To facilitate comparison with other work, we nonetheless
show the cumulative counts as well.

The differential number counts in Figure 6 have also been multiplied
by a factor of $S^{2.5}$, to remove the scaling expected for a
Euclidean universe of uniform sources; since the best-fit slope is
$-2.7$, the model is nearly flat in this representation. The solid
line shows the best fit model from the $P(D)$ analysis; the dark gray
and light gray shaded regions show the 68\% and 95\% confidence
regions, respectively. The dashed line above 4 mJy shows the
Laurent \etal counts, while the very light gray region shows the
68\% confidence region for their number counts. The agreement between
the point source-derived number counts of \citet{laur05} and the
number counts produced by the fluctuation analysis of this paper is
excellent, but the fluctuation analysis provides much tighter
constraints and extends to much lower flux densities. Also plotted
(the thick error bars at 1, 3 and 6 mJy) are the two-sided 95\%
confidence Poisson errors on the number counts; these have been
derived for a Lockman Hole-sized field and then scaled to one square
degree.

Figure 7 plots in the same fashion the cumulative number counts.  The
apparent discrepancy between the $P(D)$ results and the Laurent \etal
results as $S_\nu$ approaches 7 mJy is a consequence of the imposition
of a high-flux cutoff of 7.4 mJy in the latter analysis, in
consequence of the absence of any sources brighter than 7 mJy. To
emphasize that this apparent discrepancy is not significant, in Figure
7 we also plot the Poisson errors on the best-fit number counts for 7,
8 and 9 mJy; as for the differential number counts, these have been
calculated for a field the size of the Lockman Hole and then scaled to
one square degree. As was also evident in Figure 6, the Lockman Hole
field is simply not large enough in area to provide strong constraints
on the number density of sources at flux densities significantly in
excess of 7 mJy.

\section{Discussion and Implications}
The Bolocam Lockman Hole observations \citep{laur05} have provided
some of the first significant observational constraints on the number
counts of high-redshift galaxies at $\lambda \approx 1.1$mm. In this
paper we have taken advantage of the extremely uniform noise level of
this data set and performed a fluctuation analysis. Since it is not
necessary to extract point sources for this analysis, we are able to
probe the number count distribution to lower flux density levels than
in the previous paper, and provide substantially tighter constraints
on the slope and amplitude of the number counts at this wavelength
than any previous work.

The important results in this paper are presented in Figures 4, 5, 6
and 7. The best-fitting power-law number count model has an index
$\delta=2.7$\pma 0.18 0.15, a differential number density at 1 mJy
$n_o\simeq 1595$\pma 85 238 mJy$^{-1}$ deg$^{-2}$, and an integrated
number density $N(> 1\;{\rm mJy})\simeq 940$\pma 50 140 deg$^{-2}$
(95\% confidence limits).

At present there are few other observational results or theoretical
predictions for the 1.1mm number counts to which our results can be
compared. We have overplotted the relevant observational and
theoretical values on our derived number counts in Figures 6 and 7.

The most direct observational comparison is with the 1.2mm MAMBO
results of \citet{greve04}. We have used the source catalogs for the
Lockman Hole and ELAIS N2 fields presented in that paper, along with
the completeness and bias corrections they adopted (kindly provided by
T.~Greve) to calculate the differential counts for the combined
fields. The counts were calculated for 1 mJy-wide bins centered at
3.25, 4.25, and 5.25 mJy. The number counts and the 95\% two-sided
Poisson error bars, scaled to one square degree, are plotted as the
solid circles in Figure 6. The slopes are essentially identical, but
the normalization implied by the MAMBO number counts is higher than
our best-fit value: the MAMBO counts are higher by factors of 2.3,
2.5, and 3.3 at 3.25, 4.25, and 5.25 mJy, respectively. We show the
Lockman histogram together with the best-fitting model ($\delta\approx
3.2$) with the MAMBO 2.75 mJy normalization in Figure 8a; this model
has $\chi^2=77.3$ for 57 degrees of freedom, and hence is {\it
statistically} a much worse fit than our result. However, the
discrepancy between the two results is arguably not significant,
because the errors in the normalizations are dominated by the
systematic uncertainties in the flux densities.

The flux bias correction determined by \citet{laur05} (note that this
correction does {\it not} include the effects of Eddington bias, which
are automatically incorporated into the simulations) is
$\epsilon=0.71$\pma 0.08 0.10 ($90\%$ errors). The uncertainty on the
bias correction translates directly into a systematic
uncertainty\footnote{The quoted errors on $\epsilon$ do not take into
account any uncertainties in the flux densities of the \citet{sand94}
calibrator sources, many of which may be extended at 1.1mm.} in the
number counts, since the normalization $n_o\propto
\epsilon^{-\delta}$. For $\delta\approx 2.5$ - 3, the resulting
uncertainty in $n_o$ is approximately a factor of two. Obviously, this
systematic error term is much larger than the statistical
uncertainties on $n_o$ resulting from the $P(D)$ analysis, and is
larger than the shot noise (Poisson) errors for $S_\nu\lessapprox 3$
mJy. The true uncertainty on $n_o$ is entirely dominated by how well the
absolute calibration can be established. Similar considerations apply,
of course, to any observational determination of the number counts,
such as the \citep{greve04} results; they quote an absolute
calibration uncertainty of 20\%. At the best-fit power-law index
$\delta=2.7$, the 90\% systematic errors imply that $n_o$ could be
50\% larger, or 25\% smaller, than our quoted value. An error in the
absolute calibration affects only the normalization, $n_o$ (or $N$),
and hence will affect the vertical position and shape of the
confidence regions (since it is a $\delta$-dependent term), but not
the location or extent of these regions along the
$\delta$-axis. Hence, given the magnitude of the systematic
uncertainties in both the Bolocam and MAMBO data sets, the number
count results appear to be consistent. 

Also plotted in Figure 6 are the model differential number counts of
\citet{blain02} (open diamonds). The model predicts a slope
$\delta\simeq 3.1$ and a differential number count $n(S)=4500$
mJy$^{-1}$ deg$^{-2}$ at 1 mJy, nearly a factor of three above our
best-fit value and (formally) many $\sigma$ away from the $P(D)$
minimum; the resulting histogram is shown with the Lockman data in
Figure 8b. Although \citet{blain02} do not report uncertainties on the
model predictions, allowing for a reasonable factor of two error on
the number counts \citep{blain04} still places the \citet{blain02}
prediction more than $5\sigma$ from our best-fit result. Because of
the steeper predicted slope compared to the value we derive, the
discrepancy decreases with increasing flux density: at 7.5 mJy our
result and the \citet{blain02} prediction are within $\simeq 10\%$ of
one another. As with the comparison with the MAMBO counts, the
magnitude of the systematic flux uncertainties makes it possible to
reconcile the model normalization (within the errors) with the
data. Both the $P(D)$ results and the MAMBO data clearly suggest a
shallower slope than the \citet{blain02} prediction.

In Figure 7 we plot the same data sets and model predictions, but now
in the form of the cumulative number counts. This is potentially a
very misleading plot. Note that, while in the differential number
count plot of Figure 6 the MAMBO data points always lie above our
best-fit results from the $P(D)$ analysis (i.e., the MAMBO
normalization is higher), in the cumulative number count plot the
MAMBO number counts\footnote{There was a minor error in the calculation
of the integral number counts reported by \citet{greve04} (Greve,
priv.~comm.). We have therefore recalculated the cumulative number
counts. Only the lowest flux density bin was significantly affected,
with the result that the value for $N(> 2.75)$ mJy that we derive from
their results is about 15\% lower than the value quoted in that
paper.} decline to match the $P(D)$ results at 4.25 mJy
and drop below them (although not significantly) at higher flux
densities. 

The reason for this apparent discrepancy between the two plots lies in
the presence of an effective cut-off in the MAMBO number counts: there
are no sources (after bias correction) with fluxes that exceed 5.7
mJy, and the counts are derived simply by summing over the observed
sources (including a correction for completeness). Hence the apparent
convergence of the MAMBO counts to the Bolocam counts with increasing
flux density is illusory; it is the result of comparing the cumulative
number counts from the fluctuation analysis, which have been
calculated by extrapolating the best-fit model to arbitrarily high
fluxes, with the MAMBO counts. A fair comparison of the two would
require that we impose a cut-off on the $(PD)$ result as well,
in which case the cumulative $P(D)$ results would also decline at
higher flux densities, leading to a more or less constant ratio
between our number counts and the Greve et al counts, as is seen in
the differential number counts. This behavior is also seen for the
Lockman Hole point-source-derived number counts, for which the
maximum-likelihood fit had a cut-off of 7.4 mJy imposed (see
discussion at the end of \S 3).

We can use our best-fitting number count model to estimate the
fraction of the 1.1mm background radiation that has been resolved into
sources. Integrating from 1 to 10 mJy (the range included in the
fluctuation analysis), we obtain
\be
I_\nu=5.8\times 10^{-23}\;\intensity
\ee
which is $7\%$ of the 1.1mm background as determined by FIRAS
\citep{fix98}. This is about twice the value obtained directly from
the Bolocam Lockman Hole number counts, and half the result of
integrating the best-fitting maximum-likelihood number count model
from \citet{laur05}. If we extrapolate our best-fit result to below 1
mJy, we find that at the Bolocam one-source-per-beam level of about
0.3 mJy approximately 20\% of the 1.1mm background would be resolved;
at 0.1 mJy this would rise to 45\%.

The optimal design of future mm-wave surveys depends on precisely what
question one wishes to address. As pointed out by \citet{laur05} and
discussed in more detail above, the Bolocam Lockman Hole survey places
almost no constraints on the bright end of the number count
distribution, simply because the surveyed area was not large enough to
detect rare, bright objects. A survey aimed at probing this end of the
luminosity function should cover more area, at the cost of reduced
depth (for a reasonable amount of observing time). Such a survey is
currently being carried out with Bolocam as part of the COSMOS
survey. On the other hand, in order to study the sources that dominate
the cosmic background at 1.1mm, a deeper survey is required. As noted
above (\S 2), the analysis in this paper suggests that even at the 1 mJy
level, we are just barely touching the confusion limit, indicating
that a deeper survey would be worthwhile. As with the present data
set, a survey with very uniform noise is highly desirable for this
analysis, since it makes it possible to carry out a reliable
fluctuation analysis in addition to extraction of point sources. As
discussed above, however, systematic uncertainties in flux calibration
are likely to remain as the major source of uncertainty in determining
the number counts of sub/mm galaxies and related quantities, such as
the fraction of the millimeter background that has been resolved.

\acknowledgements We are grateful to the referee for very
helpful comments on this paper -- especially in emphasizing the
superiority of differential over integral number counts. Thomas Greve
kindly supplied information about the completeness and bias
corrections used in analysis of the MAMBO data. This work was
supported in part by NSF grants AST-0098737, AST-9980846, and
AST-0206158 and PPARC grants PPA/Y/S/2000/00101 and
PPA/G/O/2002/00015. D.~H.~acknowledges the support of a PPARC
Ph.~D.~Fellowship, S.~R.~G.~acknowledges Caltech for the
R.~A.~Millikan Fellowship, and G.~T.~L.~acknowledges NASA for GSRP
Fellowship NGT5-50384.

\appendix
\section{Effect of Optimal Filtering on the Pixel Distribution}

As noted in \S 2 of the text, since optimal filtering is a linear
operation, it will have no effect on the fluctuation analysis. To
demonstrate this, in Figure 9a we show (in black) the pixel
distribution of a simulated map; the x-axis is in mJy. The number
count model that was used has the same amplitude and slope as derived
from the P(D) analysis in this paper, but we have used a larger map
area ($512^2$ pixels, with 3 pixels per beam FWHM) to reduce the shot
noise for the sake of illustration. The noise (which is purely white
in this case, also for convenience) has an rms per pixel of 2.3
mJy. This map has not been optimally-filtered; the source contribution
is just barely visible in this representation as an excess on the
positive side of the distribution. Overplotted in red is the
theoretical P(D) distribution predicted for this model; this depends
only on the assumed number count model and the noise distribution, and
has been binned in the same way as the observed pixel distribution.

For simplicity, we have taken the optimal filter to be identical to
the beam. Figure 9b, labelled "Optimally-filtered data", plots the
same two quantities, but now for the filtered map. The signal is far more
prominent in this plot, because of the improvement in signal to noise
of the map due to the optimal filtering. The rms of the noise has been
reduced by a factor of 2.3. (We don't gain as large a reduction in the
Lockman Hole data, because of the $1/f$ noise remaining in the map
even after cleaning.) Shown in red again is the theoretical $P(D)$
distribution. This is again the predicted value, not a fit: the only
thing that has been altered in calculating this distribution is the
RMS of the noise and the number of sources per beam (since the PSF of
sources in the filtered map is larger by root two as a result of the
convolution), as well as a slight reduction in the number of pixels
(since pixels close to the map edge must be discarded when the map is
filtered). There is precise agreement between the ``observed'' and
predicted distibutions, except at the most extremal bins, where the
effects of shot noise are becoming large; this is why we discard such
bins in the analysis.



\clearpage
\setcounter{figure}{0}
\begin{figure}
\includegraphics[scale=.7]{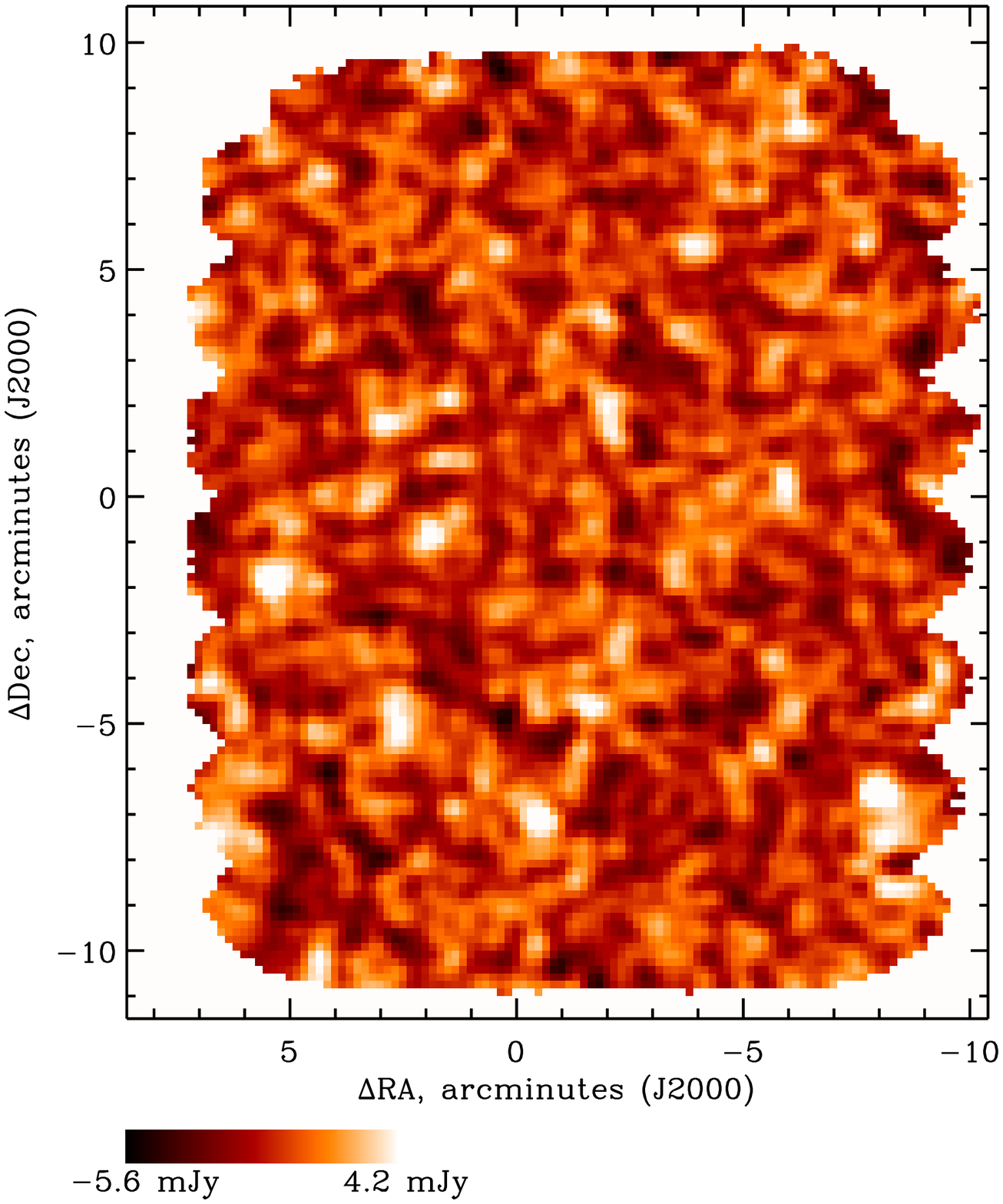}
\caption{The Bolocam 1.1mm map of the Lockman Hole region. Only the
  good coverage region is shown; the rest of the map has been masked
  off. The map is centered on RA (J2000) $10^{\rm h}52^{\rm m}08.82^{\rm
  s}$ and Dec (J2000) $+57^\circ21'33.8''$. The map pixels are
  $10''\times 10''$ and the RMS is 1.4 mJy; the color scale ranges from
  $-5.6$ to $+4.2$ mJy.}
\end{figure}

\begin{figure}
\includegraphics{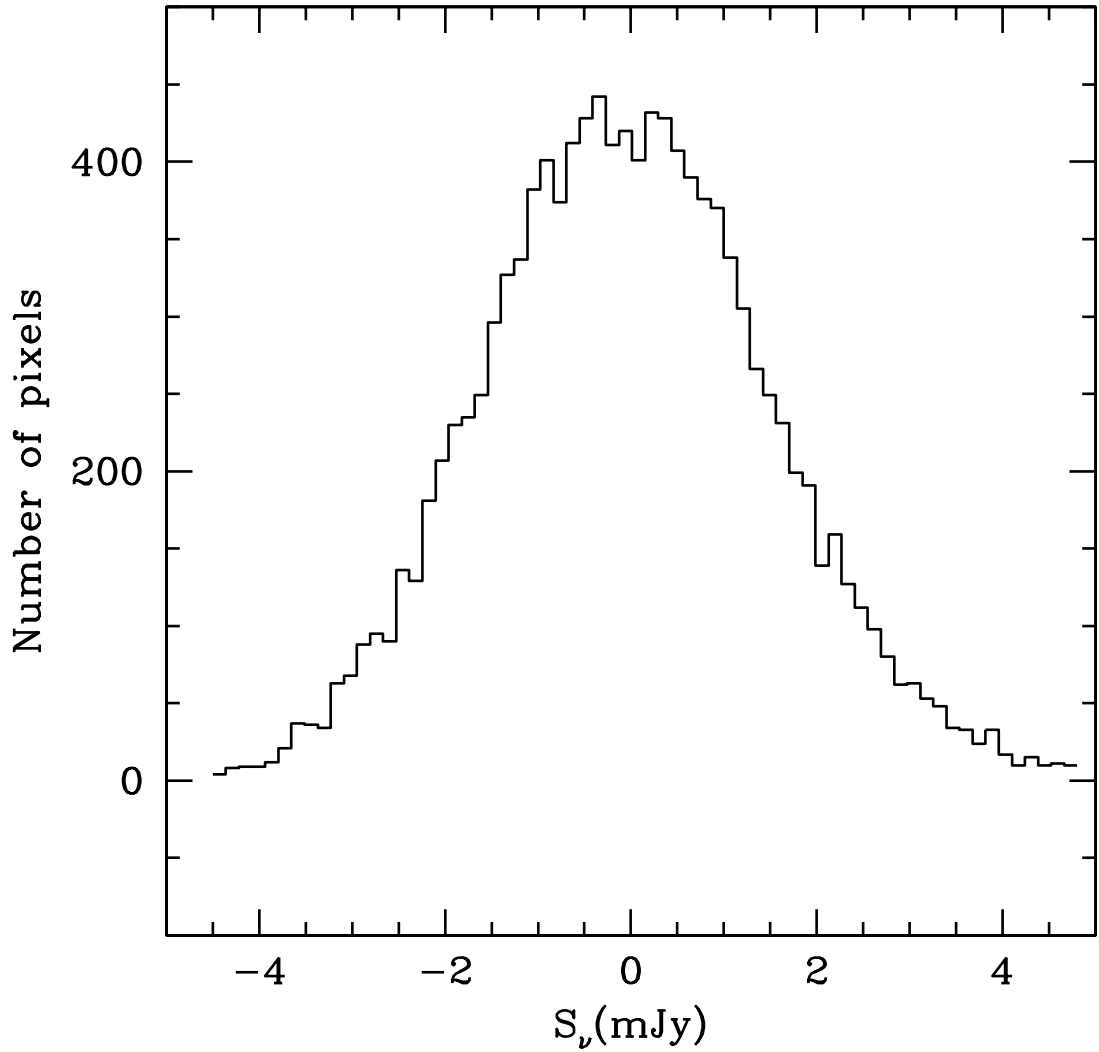}
\caption{Pixel flux density distribution of the Lockman Hole map shown
in Figure 1. There are 66 bins in flux density, extending from $-4.5$
to $+4.8$ mJy.}
\end{figure}

\begin{figure}
\includegraphics[width=8cm]{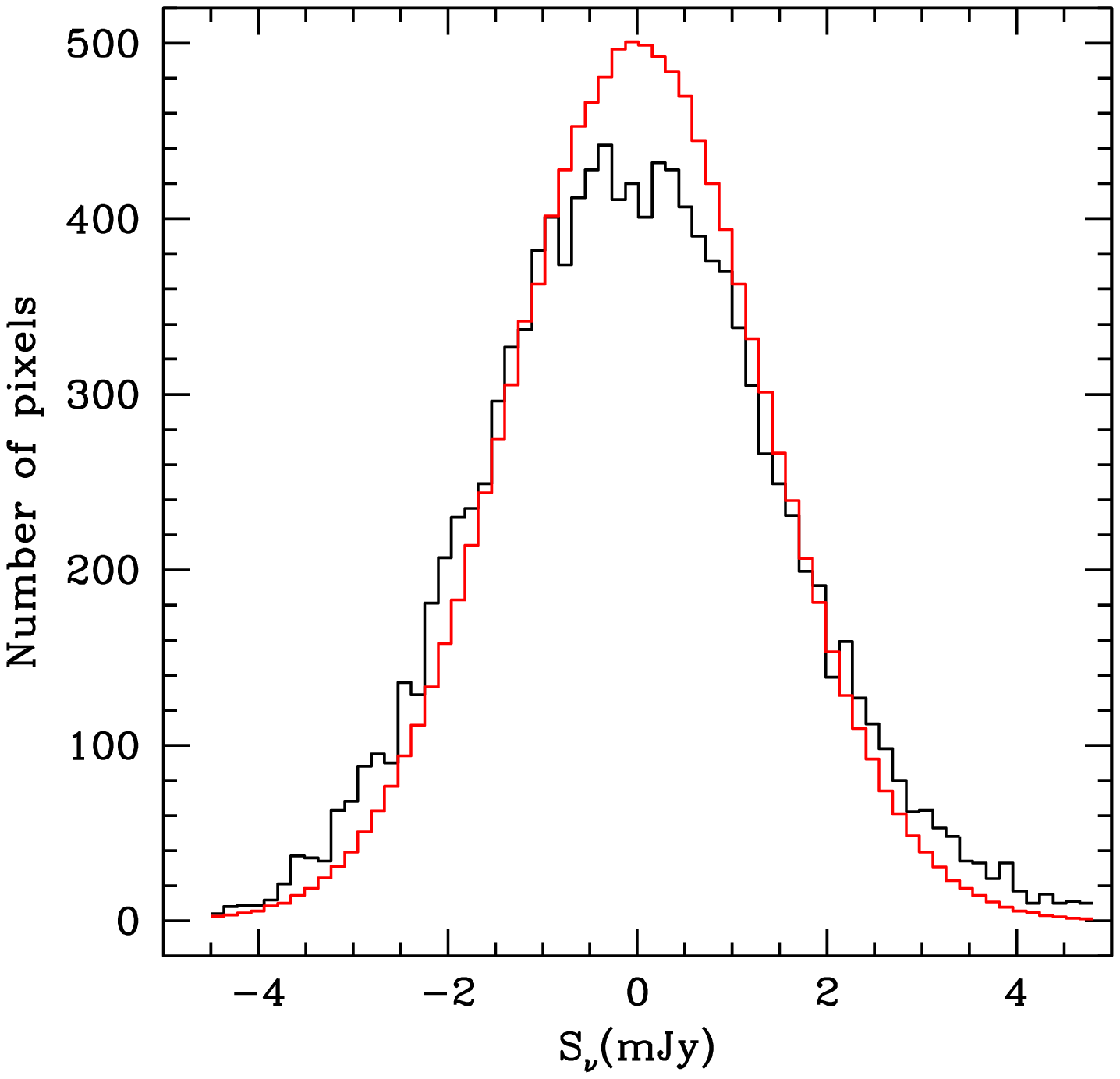}
\includegraphics[width=8cm]{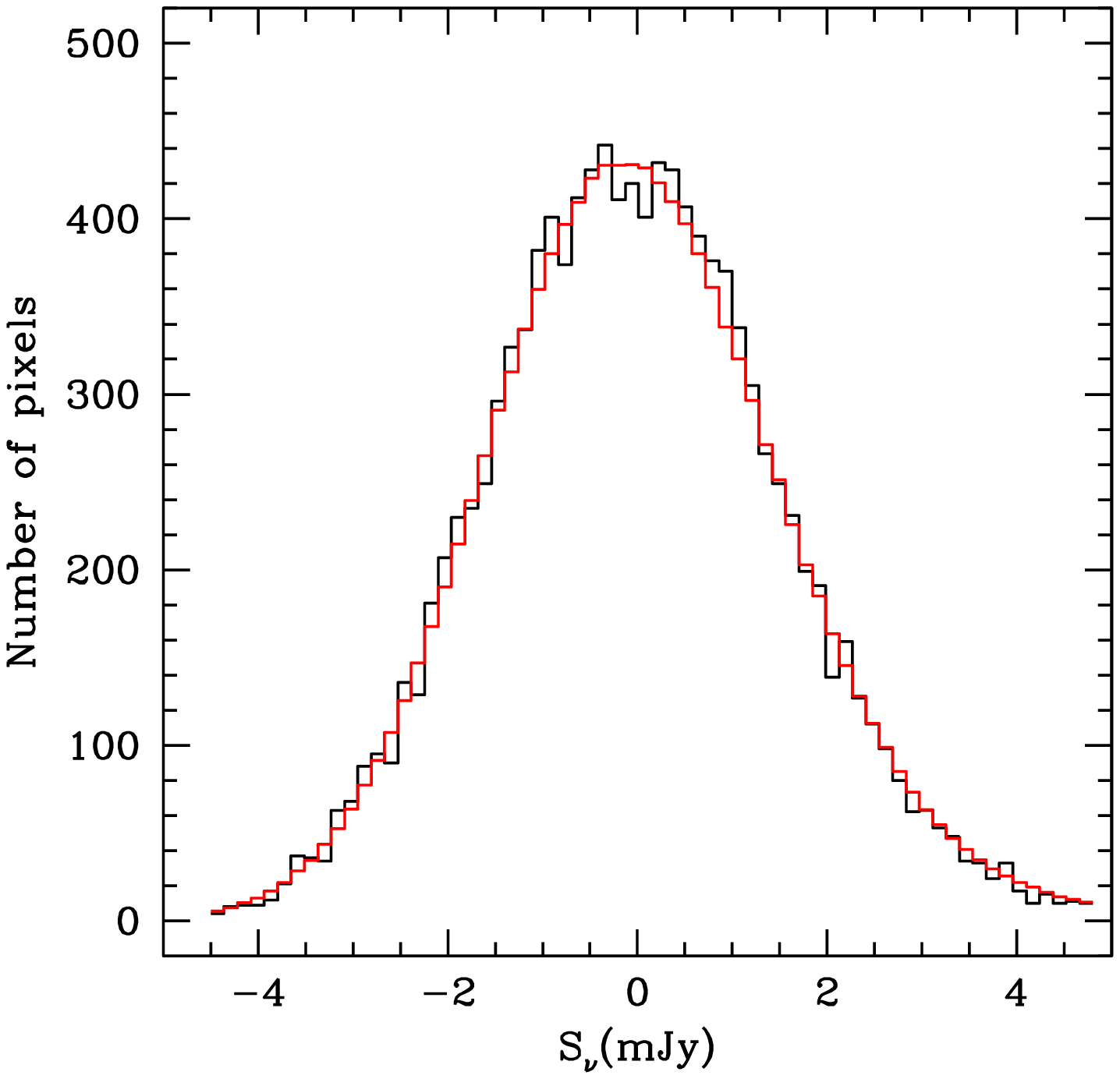}
{\small Fig.~3.~--- {\it Left:} The Lockman Hole $P(D)$ distribution
  from Figure 2 (black), overplotted with the pixel distribution
  produced by 100 realizations of a noise-only map (red). The marked
  discrepancies between the two are a consequence of the signal in the
  actual map. {\it Right:} The actual $P(D)$ as in the left panel, now
  overplotted (in red) with the theoretical $P(D)$ produced by the
  best-fitting power-law model, with $\delta=2.7$ and $n_o=1595$
  mJy$^{-1}$ deg$^{-2}$. This model has $\chi^2=51.5$, for 59 degrees
  of freedom.}
\end{figure}
\setcounter{figure}{3}

\begin{figure}
\includegraphics[width=16cm]{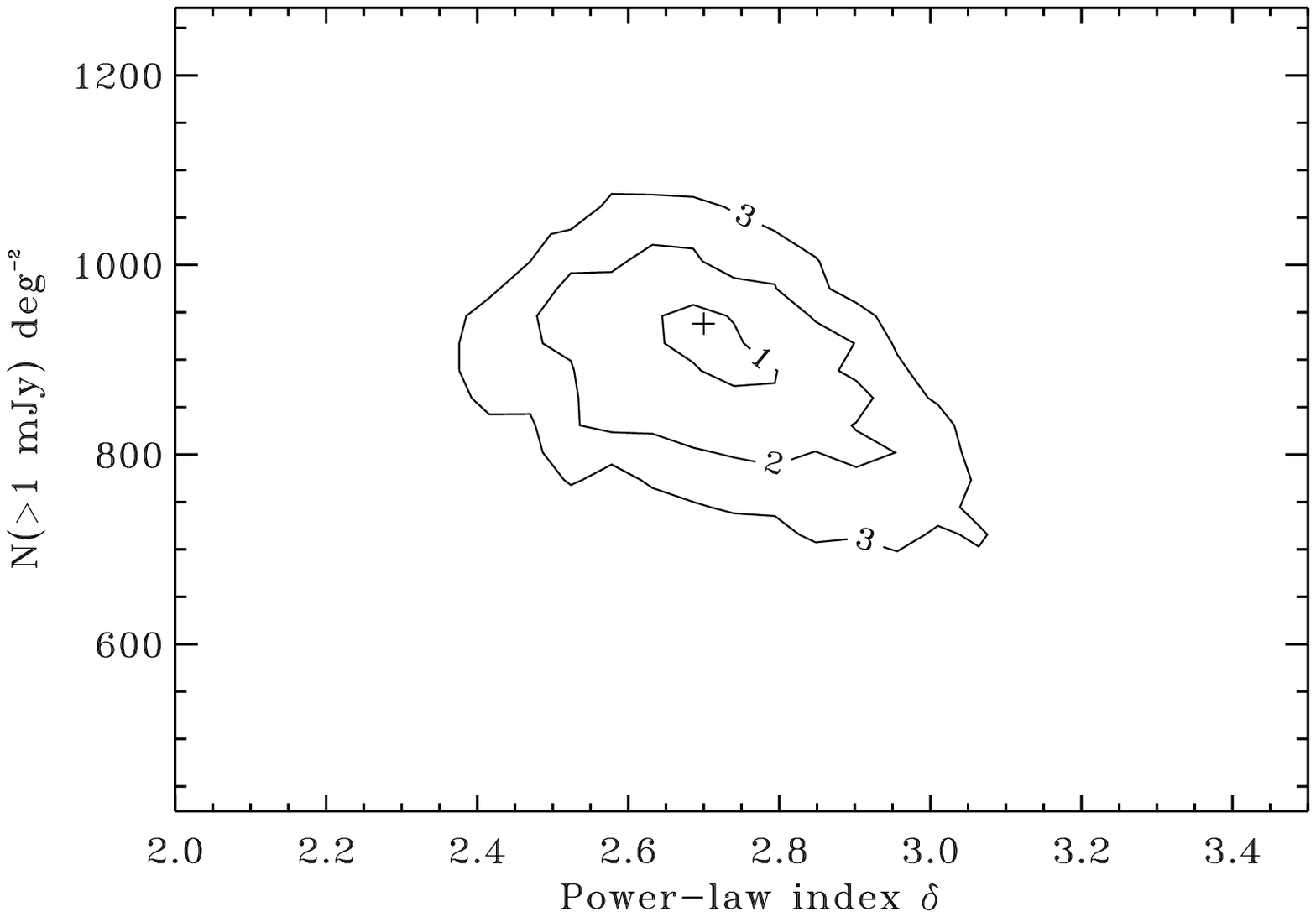}
\caption{Joint confidence limits for fit of power-law number count
  models to the Lockman Hole $P(D)$ distribution. The abscissa is the
  index of the differential power-law, while the ordinate is the
  normalization, taken to be number of sources per square degree with
  peak flux densities greater than or equal to 1 mJy. The minimum
  value of $\chi^2$, $\chi^2=51.5$, marked by a cross, is located at
  $\delta\simeq 2.7$, $N(> 1\;{\rm mJy})\simeq 938\ {\rm deg}^{-2}$;
  this corresponds to a normalization for the differential number
  counts $n_o=1595$ mJy$^{-1}$ deg$^{-2}$. The joint confidence
  regions for $\delta$ and $N$ are labeled with the values of
  $\Delta\chi^2$ as follows: $1$ for $\Delta\chi^2=2.3$, $2$ for
  $\Delta\chi^2=6.17$, and $3$ for $\Delta\chi^2=11.8$. The contours
  have been mildly smoothed by re-gridding onto a uniform grid to
  further reduce the noise.}
\end{figure}

\begin{figure}
\includegraphics[width=16cm]{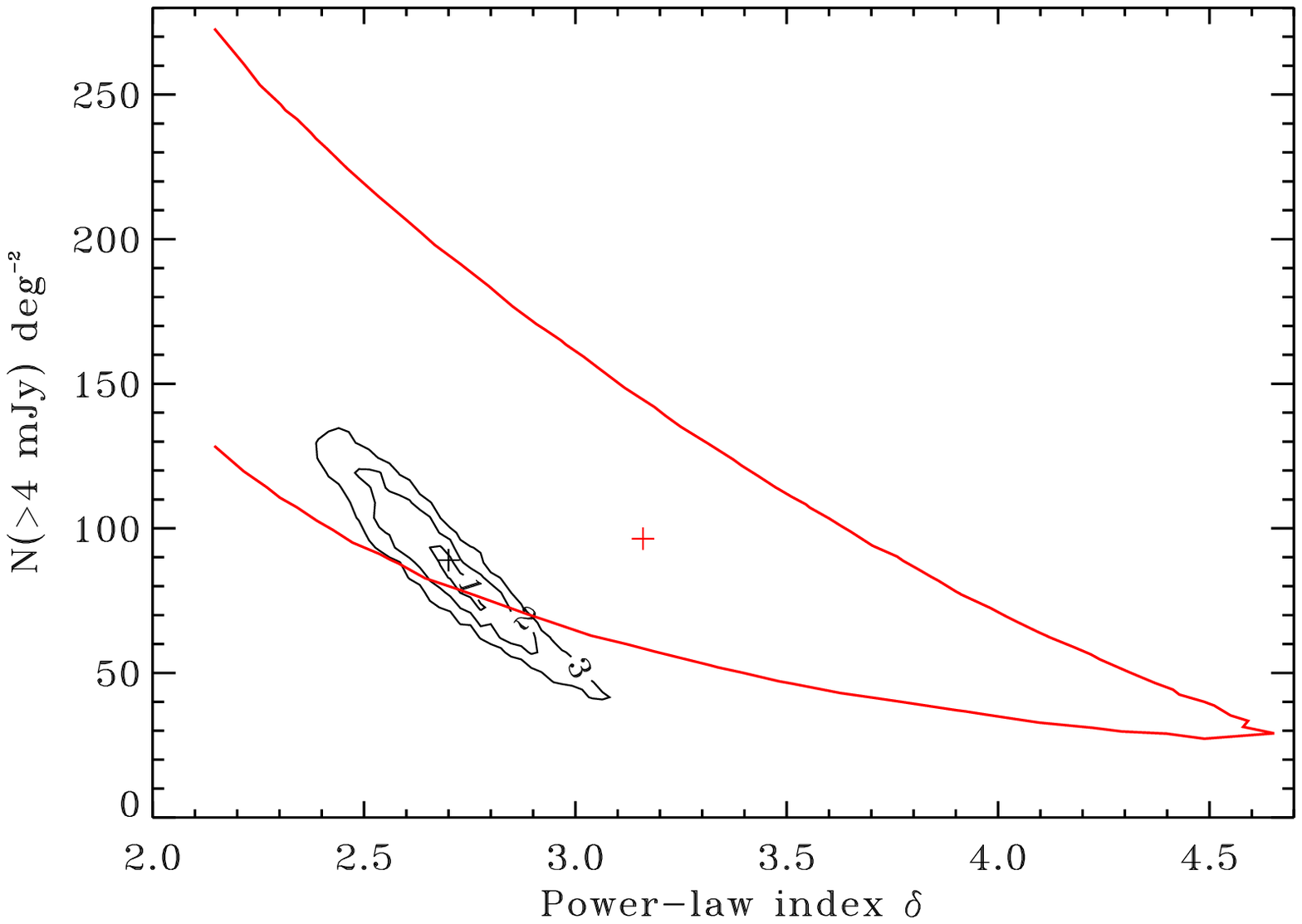}
\caption{As Figure 4, but for the integral 4 mJy counts, for
  comparison with results of \citet{laur05}. The minimum value of
  $\chi^2$ (marked by a cross) is located at $\delta\simeq 2.7$,
  $N(> 4\;{\rm mJy})\simeq 88.9\ {\rm deg}^{-2}$. The joint confidence
  regions for $\delta$ and $N$ are labelled as in Figure 4. Shown in
  red are the joint $68\%$ confidence region determined by
  \citet{laur05} (this is truncated due to the prior assumption that
  $\delta > 2$) and their derived minimum, at $\delta=3.2$ and $N(>
  4\;{\rm mJy})\simeq 96.3\ {\rm deg}^{-2}$. The agreement between the
  two methods is very good.}
\end{figure}

\begin{figure}
\includegraphics[scale=.8]{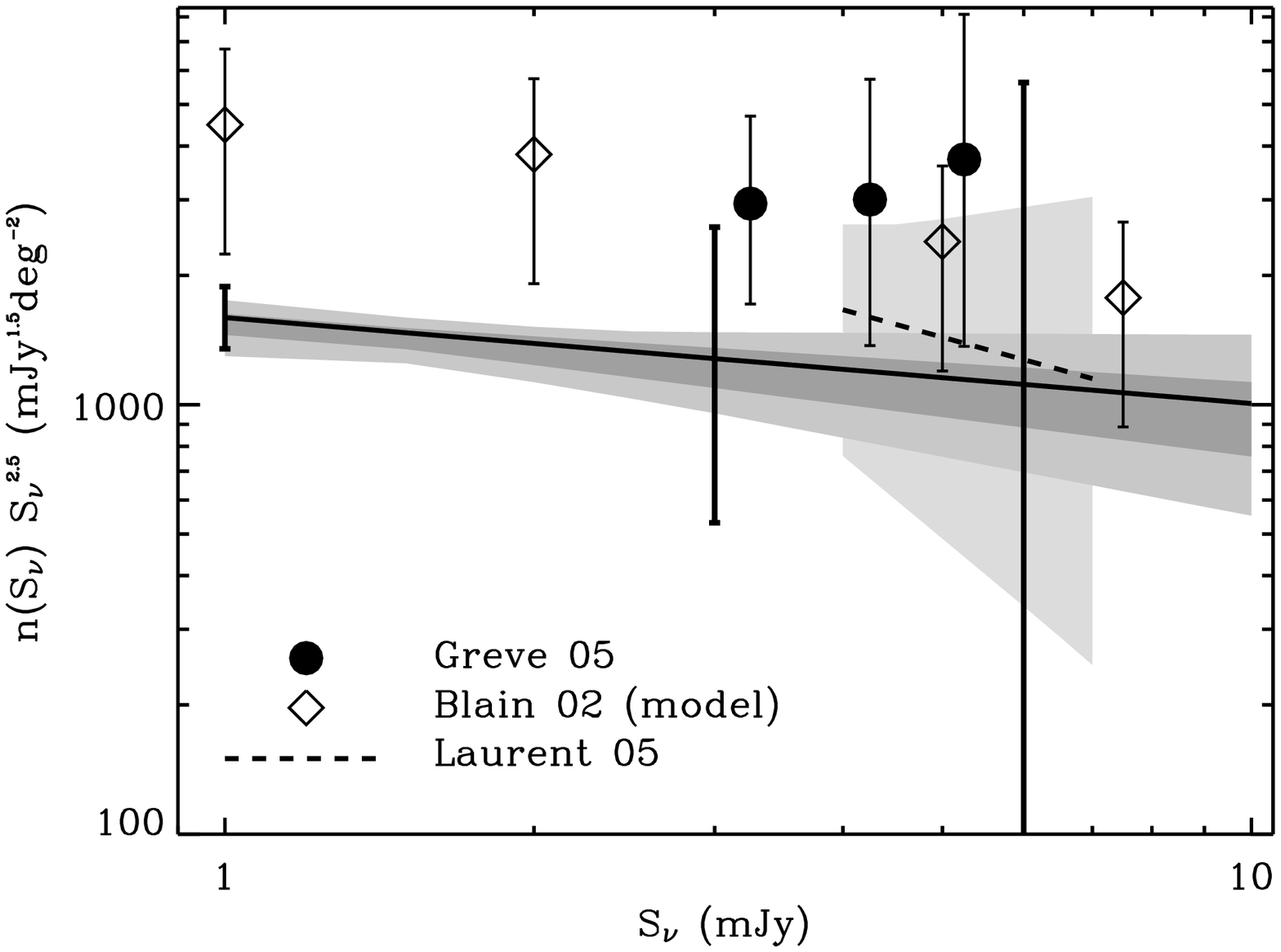}
\caption{Observed and theoretical differential number counts
  at $\lambda\simeq 1.1$mm. The counts have been scaled by
  $S^{2.5}$. The solid line shows the best-fit model 
  from the fluctuation analysis of this paper; the dark gray and light
  gray shaded regions show the 68\% and 95\% confidence limits. The
  dashed line above 4 mJy is the best-fit model from the number-count
  analysis of \citet{laur05}, while the very light gray region
  displays their 68\% confidence region. The filled circles are
  derived from the 1.2mm MAMBO observations of \citet{greve04}; the
  error bars indicate the Poisson two-sided 95\% confidence limits,
  calculated for the observed area (370 square arcminutes) and scaled
  to one square degree. The open diamonds are the model number counts
  of \citet{blain02} at 1, 2, 5, and 7.5 mJy (we have plotted only a
  few points rather than the full range simply for clarity); the error
  bars have been taken to be a factor of two \citep{blain04}. The
  thick error bars plotted on the $P(D)$ results at 1, 3 and 6 mJy
  show the two-sided 95\% confidence Poisson errors on the differential
  number counts assuming that the best-fit model is correct. These
  errors were derived for a Lockman Hole-sized field and scaled to one
  square degree.}
\end{figure}

\begin{figure}
\includegraphics[scale=.8]{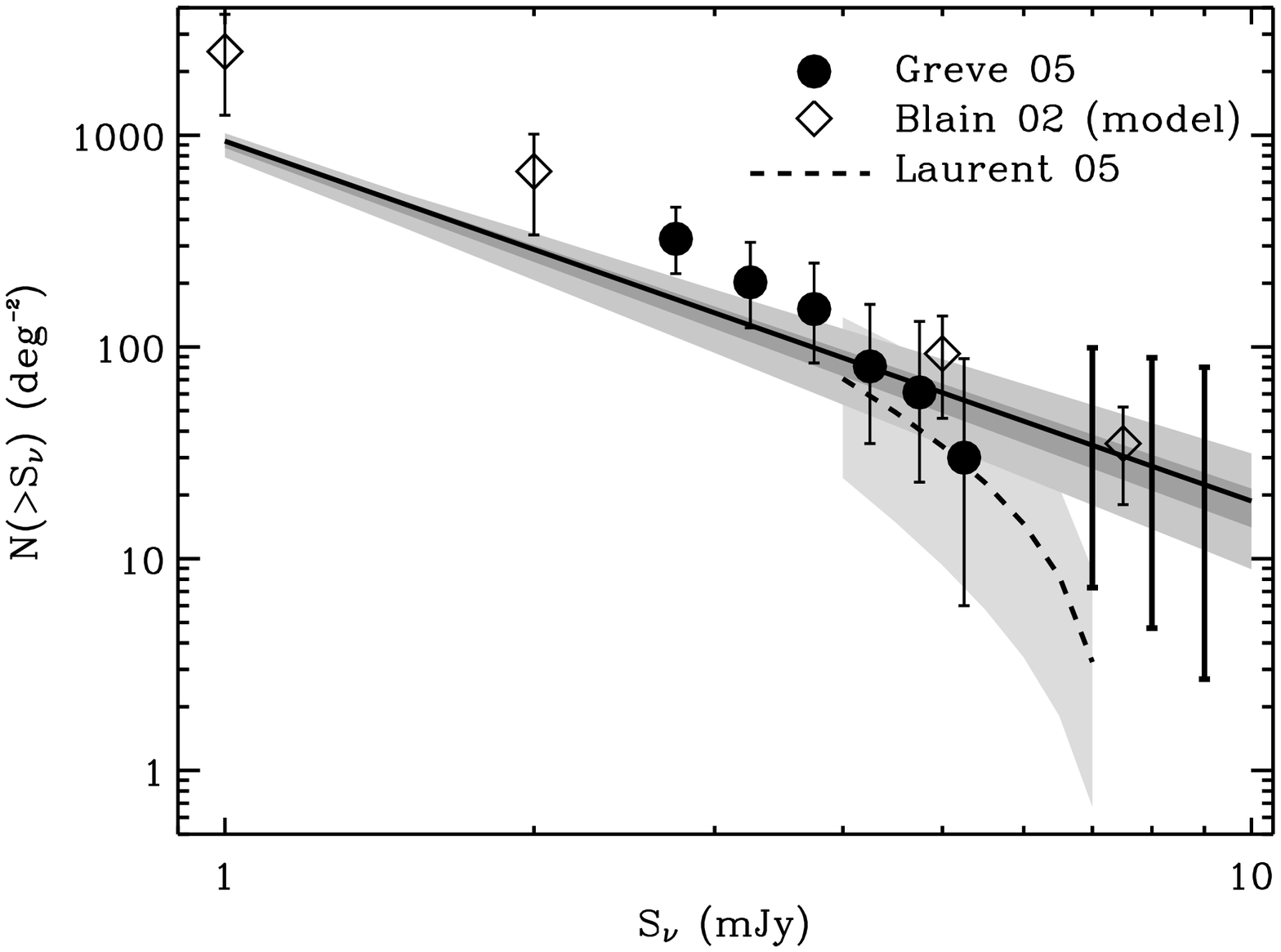}
\caption{Observed and theoretical integrated number counts
  at $\lambda\simeq 1.1$mm. The solid line shows the best-fit model
  from the fluctuation analysis of this paper; the dark gray and light
  gray shaded regions show the 68\% and 95\% confidence limits. The
  dashed line above 4 mJy is the best-fit model from the number-count
  analysis of \citet{laur05}, while the very light gray region
  displays their 68\% confidence region. The filled circles are the
  1.2mm MAMBO number counts derived by \citet{greve04}; the error bars
  are two-sided 95\% Poisson confidence limits. The open diamonds are
  the model number counts of \citet{blain02} at 1, 2, 5, and 7.5 mJy
  (we have plotted only a few points rather than the full range simply
  for clarity); the error bars have been taken to be a factor of two
  \citep{blain04}. The thick error bars plotted on the $P(D)$ results
  at 7, 8 and 9 mJy show the two-sided 95\% confidence Poisson errors
  on the integrated number counts assuming that the best-fit model is
  correct. As in Figure 6, these errors were derived for a Lockman
  Hole-sized field and scaled to one square degree.}
\end{figure}

\begin{figure}
\includegraphics[width=8cm]{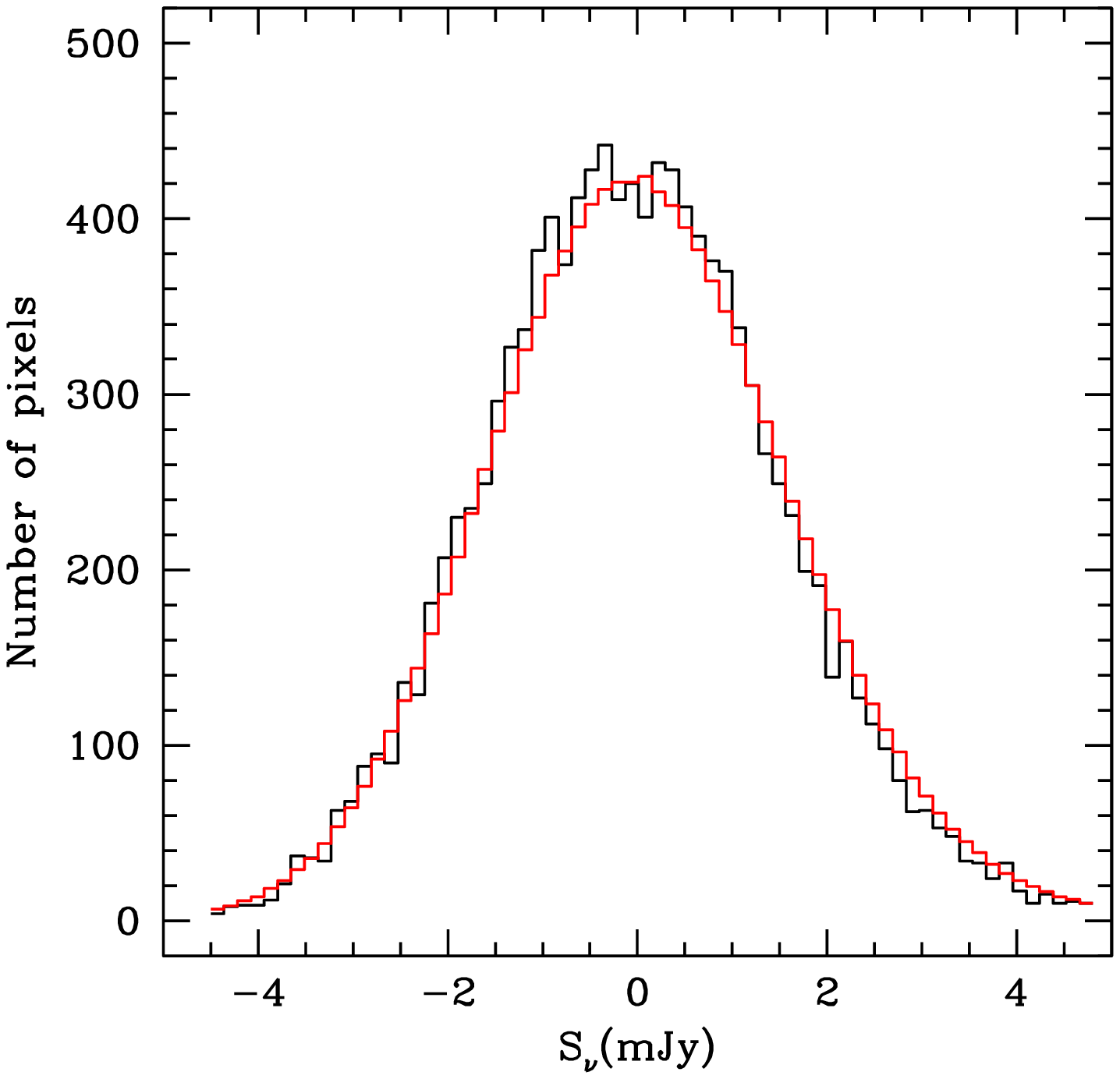}
\includegraphics[width=8cm]{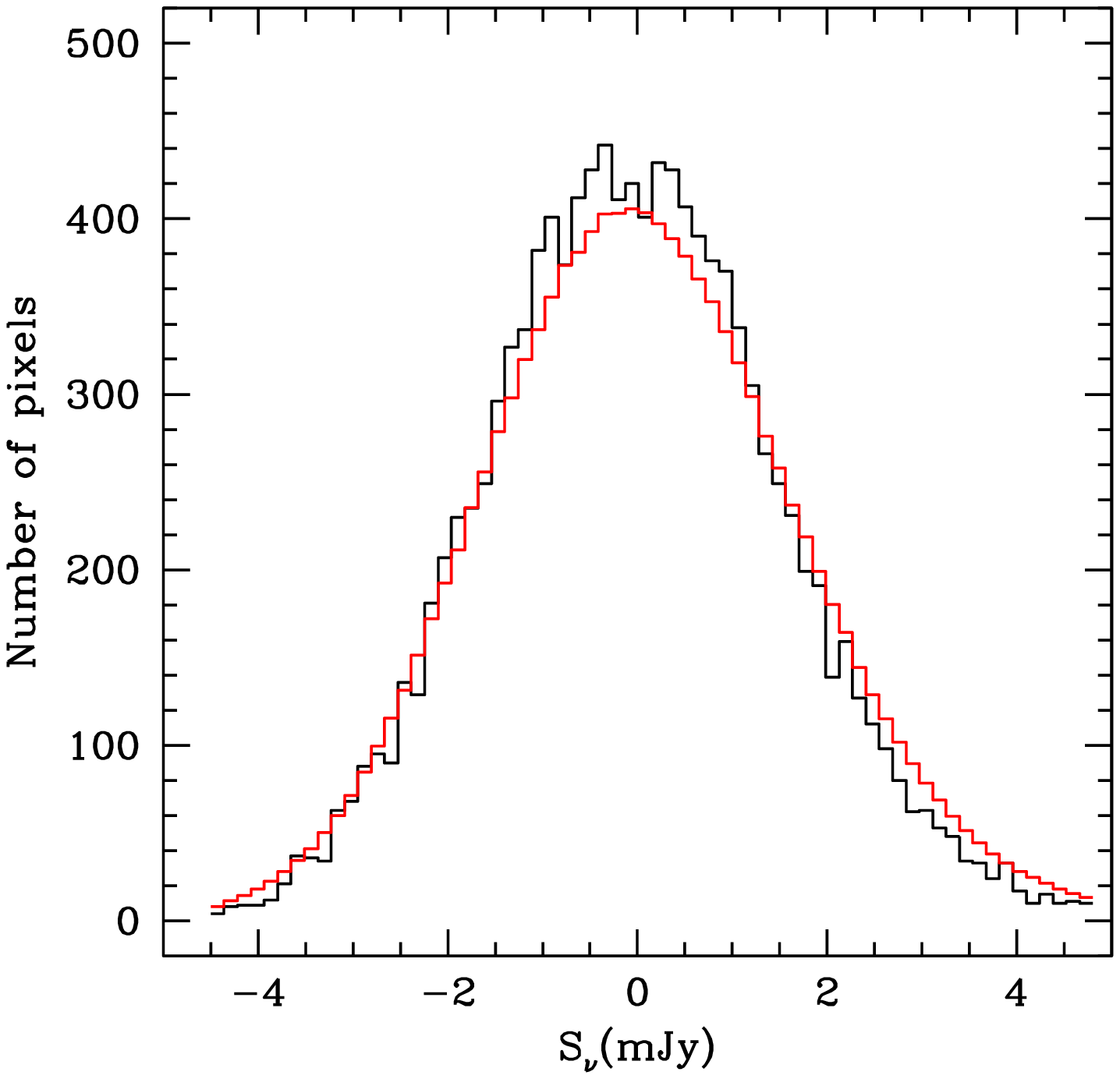}
\caption{
{\small ({\it left}): The Lockman Hole $P(D)$ distribution
  from Figure 2 (black), overplotted with the pixel distribution
  produced by 200 realizations of the best-fitting model
  ($\delta=3.2$) using the MAMBO \citep{greve04} normalization (red).
  This model has $\chi^2=77.3$ (for 57 degrees of freedom)
  compared to $\chi^2=51.5$ for the best-fitting model from the
  fluctuation analysis 
  (see Figure 3). ({\it right}): The same as Figure 8a, for
  the \citet{blain02} prediction of $\delta=3.1$, $N(> 1\;{\rm
  mJy})=2480$ deg$^{-2}$. This model has $\chi^2=133$.}
}
\end{figure}
\setcounter{figure}{8}

\begin{figure}
\includegraphics[width=8cm]{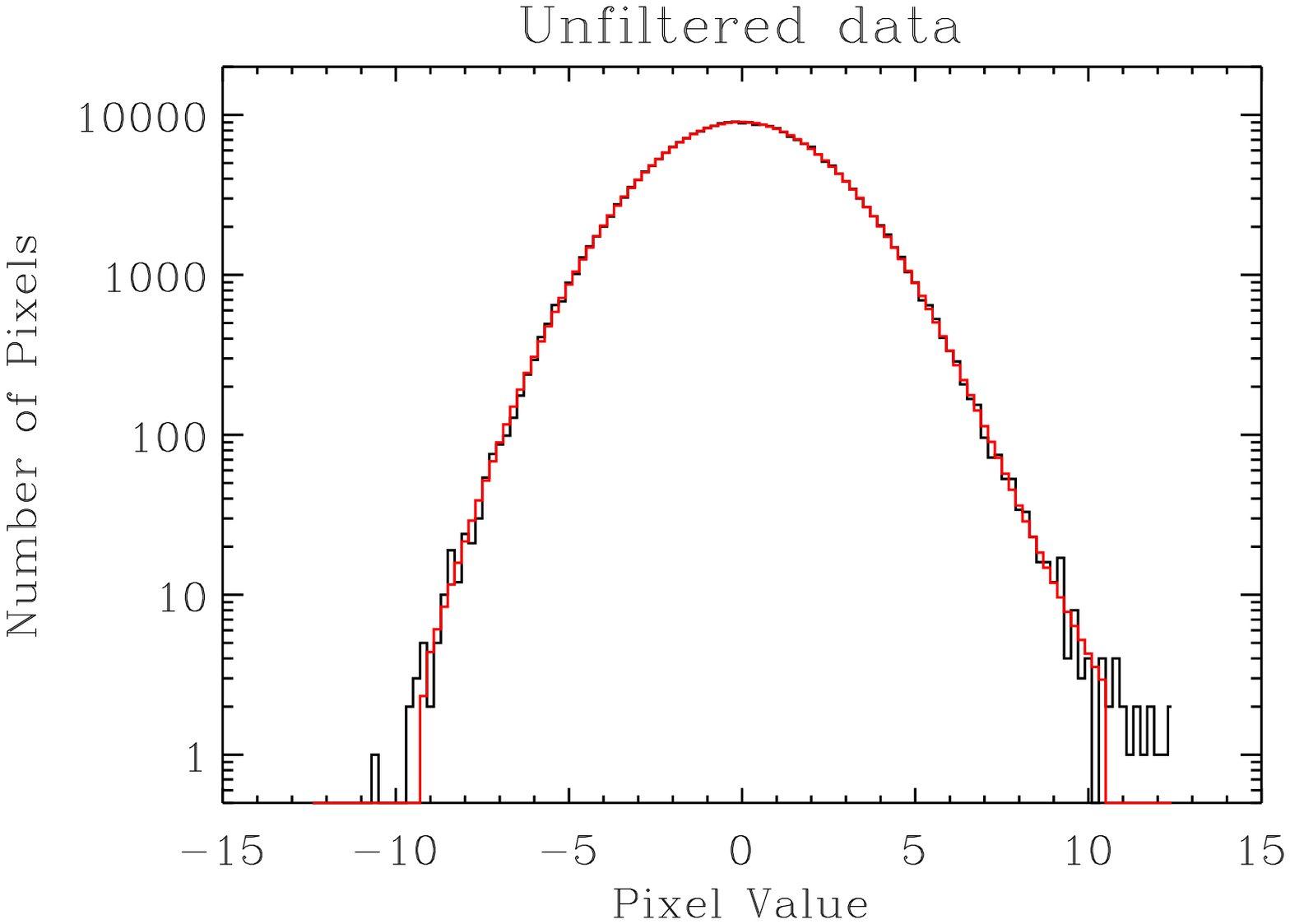}
\includegraphics[width=8cm]{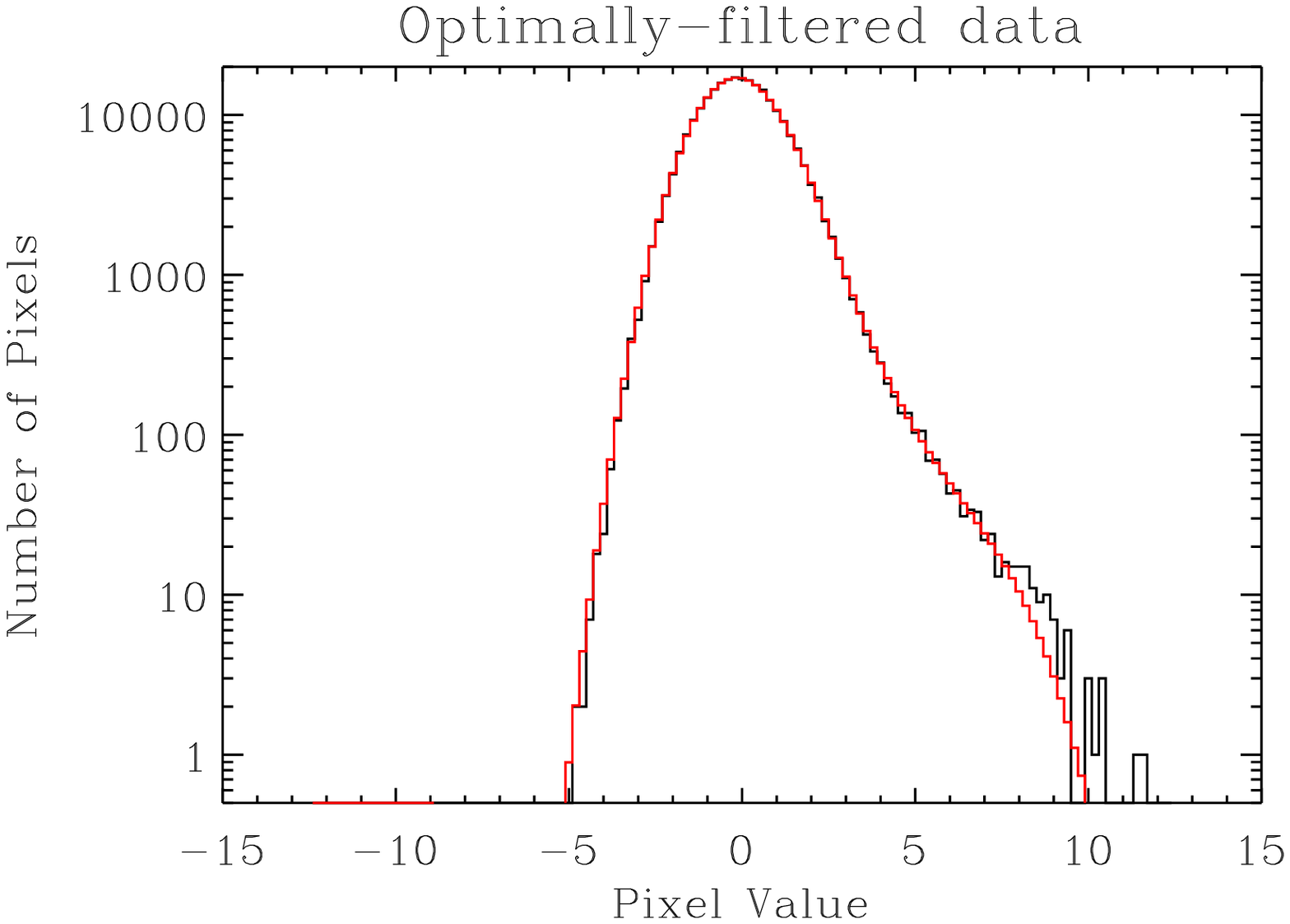}
\caption{
{\small ({\it left}): Pixel distribution of a simulated
  map (black), using the the best-fitting model from the fluctuation analysis,
  with white noise with an rms of 2.3 mJy per pixel. Overplotted in
  red is the theoretical $P(D)$ distribution; this depends only on the
  assumed number count model, the noise level, and the number of
  pixels in the map. The beam is assumed to be a Gaussian with a FWHM
  of 3 pixels, and the map is 512 pixels on a side.
  (see Figure 3). ({\it right}): The same as Figure 9a, but
  for the optimally-filtered map. In this case the optimal filter was
  assumed to be identical to the beam. The map is now 504 pixels on a
  side, since pixels close to the map boundaries must be discarded to
  avoid edge effects in filtering. The rms noise per pixel has
  been reduced to 1 mJy by filtering, with the result that the signal
  in the map is far more prominent. As in Figure 9a, the theoretical
  $P(D)$ distribution precisely matches the ``observed'' distribution,
  except for the most extremal bins, which are affected by shot noise;
  these are discarded in our analysis.}
}
\end{figure}
\setcounter{figure}{9}

\end{document}